\documentclass[twocolumn,showpacs,floatfix,superscriptaddress,amsmath,amssymb,prl]{revtex4}
\usepackage{graphicx}
\usepackage{hyperref}

\newcommand{\be}{\begin{equation}}
\newcommand{\ee}{\end{equation}}

\begin{document}

\title{Entanglement and boundary critical phenomena}

\author{Huan-Qiang Zhou}
\affiliation{Department of Mathematics, University of Queensland, Brisbane, Qld 4072, Australia}
\author{Thomas Barthel}
\affiliation{Institute for Theoretical Physics C, RWTH Aachen, D-52056 Aachen, Germany}
\author{John Ove Fj{\ae}restad}
\affiliation{Department of Physics, University of Queensland, Brisbane, Qld 4072, Australia}
\author{Ulrich Schollw\"{o}ck}
\affiliation{Institute for Theoretical Physics C, RWTH Aachen, D-52056 Aachen, Germany}

\begin{abstract}
We investigate boundary critical phenomena from a quantum
information perspective. Bipartite entanglement in the ground state
of one-dimensional quantum systems is quantified using the R\'{e}nyi
entropy $S_{\alpha}$, which includes the von Neumann entropy
($\alpha \to 1$) and the single-copy entanglement
($\alpha\to\infty$) as special cases. We identify the contribution
from the boundary entropy to the R\'{e}nyi entropy, and show that
there is an entanglement loss along boundary renormalization group
(RG) flows. This property, which is intimately related to the
Affleck-Ludwig $g$-theorem, can be regarded as a consequence of
majorization relations between the spectra of the reduced density
matrix along the boundary RG flows. We also point out that the bulk
contribution to the single-copy entanglement is \textit{half} of
that to the von Neumann entropy, whereas the boundary contribution
is \textit{the same}.
\end{abstract}

\date{\today}

\maketitle

Recently much work has been done to understand entanglement in
quantum many-body systems. In particular, the behavior of various
entanglement measures at or near a quantum phase transition
\cite{sachdev} has received a lot of attention
\cite{nielsen,latorrevidal,korepin,levine,calabrese,casini1,casini2,eisert}.
These entanglement measures include the von Neumann entropy and the
single-copy entanglement, among others. The former is the most
studied measure and quantifies entanglement in a bipartite system in
the so-called asymptotic regime \cite{bennett}, whereas the latter
was recently suggested to quantify the entanglement present in a
single copy \cite{eisert}. For a system in a pure state
$|\psi\rangle$ (e.g.\ the ground state) that is partitioned into two
subsystems $A$ and $B$, the von Neumann entropy is $S_1\equiv
-\mbox{Tr}_A \rho_A \log_2 \rho_A$ where $\rho_A=\mbox{Tr}_B
|\psi\rangle\langle\psi|$ is the reduced density matrix for $A$, and
the single-copy entanglement is $S_{\infty}\equiv -\log_2
\lambda_1$, where $\lambda_1$ is the largest eigenvalue of $\rho_A$.

Studies of the von Neumann entropy for quantum spin chains
\cite{latorrevidal,korepin,levine,calabrese,casini1,casini2} have
revealed that its dependence on the size $\ell$ of the block $A$ is
very different for noncritical and critical systems. For the former,
the von Neumann entropy increases logarithmically with $\ell$ until
it saturates when $\ell$ becomes of order the correlation length
$\xi$, while for the latter (having $\xi=\infty$) it diverges
logarithmically with $\ell$. Remarkably, the prefactor of the
logarithmic term is universal and proportional to the central charge
of the underlying conformal field theory
(CFT)~\cite{wilczek,calabrese}. Furthermore, it has been shown
numerically \cite{latorre} that the \textit{entanglement loss} along
the (bulk) renormalization group (RG) flows, which is consistent
with the CFT predictions for the von Neumann
entropy~\cite{calabrese,wilczek} and with Zamolodchikov's
$c$-theorem \cite{zamolodchikov}, can be given a more
``fine-grained'' characterization in terms of the
\textit{majorization} concept \cite{olkin}. A theoretical analysis
of majorization in these systems also appeared recently \cite{orus}.

Boundary critical phenomena \cite{binder} in one-dimensional (1D)
quantum systems (equivalently, 2D classical systems) have attracted
a lot of interest, especially in the context of boundary CFT. A
closely related subject is the theory of boundary perturbations of
certain conformally invariant theories, so-called integrable
boundary quantum field theory \cite{ghoshal}, which is relevant to
quantum spin chains with nontrivial boundary interactions,
impurities in Luttinger liquids, Kondo physics, tunneling in
fractional quantum Hall devices, and open string theory. In these
problems, the Affleck-Ludwig $g$-theorem \cite{affleck} plays a
similar role as the $c$-theorem does for bulk critical phenomena.

In this Letter we investigate boundary critical phenomena in quantum
spin chains from a quantum information perspective. Our findings
provide new insights into the information-theoretic explanation of
the boundary entropy and the $g$-theorem characterizing the
intrinsic irreversibility due to information loss along boundary RG
flows \cite{preskill}. We choose the R\'{e}nyi entropy (also known
as the $\alpha$-entropy) $S_\alpha=(1-\alpha)^{-1}\log_2 \mbox{Tr}_A
\rho^\alpha_A$ as our entanglement measure, partly motivated by the
fact that both the von Neumann entropy and the single-copy
entanglement are special cases of the R\'{e}nyi entropy,
corresponding to $\alpha \to 1$ and $\alpha \rightarrow \infty$,
respectively. Using CFT we derive expressions for the R\'{e}nyi entropy
which include a boundary entropy contribution, and show that a
majorization relation underlies an entanglement loss along boundary
RG flows (i.e., when a system with a boundary interaction flows from
an unstable to a stable fixed point). We support our analytical
arguments with numerical density-matrix renormalization group (DMRG)
calculations \cite{White1992-11, Schollwoeck2005}.

\textit{R\'{e}nyi entropy and conformal field theory.}---Consider a
1D lattice of interacting spins with lattice spacing $a$. Let $L$ be
the total length of the system, and let the two subsystems $A$ and $B$
be blocks of consecutive spins of length $\ell$ and $L-\ell$,
respectively. Then CFT predicts that for an infinite spin chain at
criticality,
\begin{equation}
S_{\alpha} = \frac{c}{6}\big(1+\alpha^{-1}\big)\log_2 \frac{\ell}{a}
+c'_{\alpha}
\label{renyi-1}
\end{equation}
where $c$ is the central charge and  $c'_{\alpha}$ is a
non-universal constant \cite{calabrese}. For a semi-infinite spin
chain with the block $A$ starting at the origin where a boundary
interaction is applied, we have instead
\begin{equation}
S_{\alpha} = \frac{c}{12}\big(1+\alpha^{-1}\big)\log_2
\frac{2\ell}{a} + \frac{1}{2}c'_{\alpha} + S_b. \label{renyi-2}
\end{equation}
Here $S_b = \log_2 g$ is the boundary entropy \cite{affleck}, with
$g=\langle B|0\rangle$, where $|B\rangle$ is a boundary state
\cite{cardy-lewellen,difrancesco,bs} and $|0\rangle$ is the ground
state. We emphasize that, compared with the corresponding expression
for $S_1$ in Ref.~\cite{calabrese}, there is an extra factor $1/2$
in front of $c'_\alpha$ and $S_b$ in Eq.~(\ref{renyi-2}).

The corresponding R\'{e}nyi entropies for finite $L$ are found from
Eqs. (\ref{renyi-1}) and (\ref{renyi-2}) by standard conformal
mappings \cite{difrancesco}. As a result, the R\'{e}nyi entropy for
periodic boundary conditions (PBC) [open boundary conditions (OBC)]
is given by replacing $\ell/a \to (L/\pi a)\sin \pi\ell/L$ in Eq.~(\ref{renyi-1}) [Eq.~(\ref{renyi-2})]. For the OBC case we have
assumed that the BC on the left and right ends are identical;
otherwise one has to consider the complicating effects of a boundary
condition changing operator.

From these results one sees that the contribution of the
\textit{bulk} universal part to the single-copy entanglement
$S_{\infty}$ is always {\it half} of that to the von Neumann entropy
$S_1$, thus extending the conclusion in Ref.~\cite{eisert} for the
XX chain ($c=1$) to all conformally invariant critical systems.
However, we stress that the contribution of the boundary entropy to
$S_{\alpha}$ does not depend on $\alpha$.

The details of the CFT derivation of the above results (which makes
use of CFT expressions for $\mbox{Tr}_A \rho_A^{\alpha}$ from Ref.~\cite{calabrese}) will be given elsewhere. Here we will instead
present a heuristic argument to justify Eqs. (\ref{renyi-1}) and
(\ref{renyi-2}). As is well known, the central charge $c$, which
measures the (effective) number of gapless excitations, describes
the way a specific system reacts to the introduction of a
macroscopic length scale into the system \cite{difrancesco}.
Therefore, when the entire (infinite) system is partitioned into a
block of length $\ell$ and its environment, one may expect that the
R\'{e}nyi entropy $S_\alpha$ depends only on $c$ and $\ell/a$ with
some short distance cutoff $a$, e.g.\ the lattice spacing for
quantum spin chains. Because both the R\'{e}nyi entropy and the
central charge are \textit{additive}, one can see~\cite{uni} that
the R\'{e}nyi entropy must be linear as a function of the central
charge, \textit{i.e.}, $S_\alpha = c f_\alpha(\ell/a) + h_\alpha$,
with $f_\alpha$ a universal function and $h_\alpha$ a non-universal
(\textit{i.e.}, model-dependent) function. The specific form of the
function $f_\alpha$ may be determined by calculating the R\'{e}nyi
entropy for any exactly solvable model, for instance the massless
Dirac fermion field as done in Ref.~\cite{casini2} for PBC, thus
confirming Eq.~(\ref{renyi-1}). The calculation may be extended to
the massless Dirac fermion with a conformally invariant boundary,
leading to the conclusion that $h_\alpha$ consists of the
non-universal bulk part $1/2 \; c'_\alpha$ and the universal
boundary part $S_b$. The bulk universal part of Eq.~(\ref{renyi-2})
and its counterpart for finite $L$ result from the substitution
$c\to c/2$, $\ell\to 2\ell$ and $L\to 2L$ in Eq.~(\ref{renyi-1}) and
its corresponding counterpart, due to the fact that one may
``unfold'' the system by identifying left-movers at position $x$
($x=ja$, with $j$ labeling lattice sites) with right-movers at $-x$
so that the resulting system consists of only right movers subject
to PBC and with the system size scales $\ell$ and $L$ doubled and
the number of gapless degrees of freedom halved. The factor $1/2$ in
front of $c'_\alpha$ in the second bulk term in Eq.~(\ref{renyi-2})
also originates from this halving of the number of gapless degrees
of freedom.

\textit{$S=1/2$ transverse Ising and XXZ chains in boundary magnetic
fields.}---We now address the implications of these CFT predictions
for models of semi-infinite quantum spin chains. We first consider
the $S=1/2$ transverse Ising chain in a boundary magnetic field,
described by
\begin{equation}
 H_{\textrm{Ising}} = -\sum_{j=0}^\infty
(S^x_j S^x_{j+1} + h S^z_j)+h_b S^x_0 .
\label{eq:HamIsing}
\end{equation}
Here $h$ is the transverse bulk magnetic field, and $h_b$ is the
boundary magnetic field.  We set $h=1/2$ so that the model is bulk
critical with central charge $c=1/2$. The second model is the
$S=1/2$ XXZ chain, for which
\begin{equation}
 H_{\textrm{XXZ}} = \sum_{j=0}^\infty (S^x_j S^x_{j+1}+
S^y_j S^y_{j+1} + \Delta S^z_j S^z_{j+1}) + h_b S^x_0.
\end{equation}
Here $\Delta$ denotes the anisotropy and $h_b$ is the (transverse)
boundary magnetic field. The model is bulk critical with central
charge $c=1$ for $-1< \Delta \leq 1$.

For both models, the points $h_b=0$ and $h_b=\pm \infty$,
corresponding to the free and fixed conformally invariant BC, are
boundary critical fixed points; the former is unstable and the
latter is stable. A boundary magnetic field $h_b>0$ is generally a
relevant perturbation of the free BC $h_b=0$, and generates a
boundary RG flow of $h_b$ towards the fixed point $h_b=\infty$. For
$0<h_b<\infty$, the conformal invariance is lost, and the
competition between boundary ordering and bulk criticality
introduces a crossover length $\zeta \propto h_b^{d-1}$
\cite{aff98}, where $d<1$ is the scaling dimension of the relevant
boundary perturbation. For the Ising model, $d=1/2$, and for the XXZ
model, $d=2\pi R^2$, where $R=\sqrt{(1/2\pi)-(1/2\pi^2)\arccos
\Delta}$ is the compactification radius. Furthermore, for both
models, $g$ for fixed BC is less than $g$ for free BC, which implies
that the R\'{e}nyi entropy is less for fixed than for free BC. This
is also consistent with the $g$-theorem \cite{affleck}, which states
that $g$ decreases along boundary RG flows. For the transverse Ising
model, $g=1$ (free) and $g=1/\sqrt{2}$ (fixed)
\cite{cardy-lewellen}; for the XXZ model, $g=\pi^{-1/4}(2R)^{-1/2}$
(free) and $g=\pi^{1/4}R^{1/2}$ (fixed) \cite{aff98}. We emphasize
that even away from boundary critical points, Eq.~(\ref{renyi-2}) is
still valid for $\ell > \zeta$, due to bulk criticality.

$H_{\textrm{XXZ}}$ reduces to the isotropic XXX (Heisenberg) model
for $\Delta=1$ ($R=1/\sqrt{2\pi}$). This case is special, because
the boundary perturbation is marginal ($d=1$). In fact, $g=2^{-1/4}$
for both free and fixed BC. The line from $h_b=0$ to $h_b=\infty$ is
a line of fixed points; there is no RG flow since $g$ is the same
everywhere along the line. To see this, we consider the finite XXX
chain with identical boundary interactions at the two boundaries.
This corresponds to a free scalar boson field with a dynamical
boundary interaction \cite{callan}, for which it was found that when
both ends have the same BC, the partition function is independent of
the boundary interaction strength, reflecting the fact that the
energy levels do not feel the presence of the boundary interaction.
Combining this with the CFT prediction \cite{cardy05} $S= (\pi
c/3\beta)L + 2 S_b$ for the Gibbs entropy $S\equiv -\beta^2
\partial F/\partial \beta$, we conclude that the boundary entropy does not depend on the
boundary magnetic field $h_b$.

\begin{figure}
 \centering
\includegraphics[width=7.55cm,height=6.0cm]{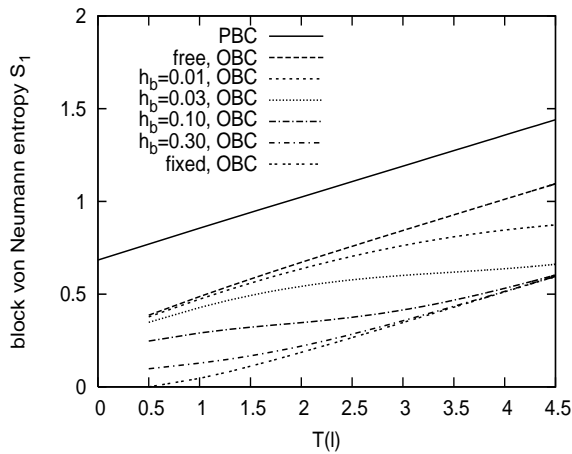}
\caption{\label{fig:IsingS1}Bipartite entanglement for the bulk
critical transverse Ising model, quantified by the von Neumann
entropy $S_1$ of the reduced density matrix. The block size $\ell$
is given as $T(\ell)$ (see text)  which allows for a direct check of
the CFT predictions (\ref{renyi-1}) and (\ref{renyi-2}). The fit
yields the predicted boundary entropies.}
\end{figure}
To check the CFT predictions, we analyze quasi-exact DMRG results.
Fig.~\ref{fig:IsingS1} shows the entanglement in the bulk critical
transverse Ising model \eqref{eq:HamIsing} for PBC and OBC with
different boundary fields $h_b$. For PBC we use the DMRG variant
described in Ref.~\cite{Verstraete2004-4}. The block size $\ell$ is
shown as $T(\ell)$, with
$T_\textrm{PBC}(\ell)=\log_2\left(\frac{L}{\pi
a}\sin\frac{\pi\ell}{L}\right)$ and
$T_\textrm{OBC}(\ell)=\frac{1}{2}\,\log_2\left(\frac{2L}{\pi
a}\sin\frac{\pi\ell}{L}\right)$, appropriate for a linear fit of the
finite $L$ counterparts of \eqref{renyi-1} and \eqref{renyi-2}. The
fits for PBC, free OBC and fixed OBC yield the desired central
charge ($c=0.500,\, 0.502,\, 0.498\approx{1}/{2}$) and the predicted
boundary entropies ($S_\textrm{b}^\textrm{free}=-0.003\approx 0$,
$S_\textrm{b}^\textrm{fixed}=-0.497\approx \log_2{1}/{\sqrt{2}}$),
which can be read off from the differences of the axis intercepts.
The curves for nonzero $h_b$ converge for large $\ell$ to the fixed
BC curve (the smaller the value of $h_b$, the larger the crossover
length $\zeta$).  This implies that the boundary entropy for any
nonzero $h_b$ is the same as that for fixed BC, consistent with the
explicit construction of boundary states in Ref.~\cite{ghoshal}.

\textit{Majorization relations.}---Let us first recall the
definition of majorization \cite{olkin}. Consider two probability
distributions $\lambda\equiv \{\lambda_i\}$ and $\mu\equiv
\{\mu_i\}$ whose elements are ordered such that $\lambda_1\geq
\lambda_2\geq \cdots \geq \lambda_r$, and similarly for $\mu$. We
say that $\lambda$ is majorized by $\mu$, written $\lambda \prec
\mu$, if $\sum _{i=1}^{k} \lambda _i \leq \sum _{i=1}^{k} \mu _i$
for $k=1,\ldots,r-1$ and $\sum _{i=1}^{r} \lambda _i = \sum
_{i=1}^{r} \mu _i$ ($=1$). Here the probability distributions are
formed by the eigenvalues of the reduced density matrix $\rho_A$,
and $r$ is the Schmidt rank. An elementary result \cite{olkin}
states that $\lambda \prec \mu$ if and only if $\phi(\lambda) \geq
\phi(\mu)$ for all Schur-concave functions $\phi$ \cite{schur}. In
fact, the R\'{e}nyi entropy is Schur-concave for any index $\alpha$.
Therefore it is necessary (but not sufficient) for the R\'{e}nyi
entropy for all indices $\alpha$ to be monotonic with some system
parameter ($\ell$ or $h_b$) for the corresponding spectra of the
reduced density matrix to be subject to a majorization relation.

The CFT results \eqref{renyi-1} and \eqref{renyi-2} show that at (both bulk and boundary) criticality
the R\'{e}nyi entropy increases monotonically with the block size
$\ell$ (for this to hold for a finite chain of length $L$, $\ell$
must be less than $L/2$). In particular, the largest eigenvalue
$\lambda_1=2^{-S_{\infty}}$ of $\rho_A$ decreases with increasing
$\ell$. This indicates that $\rho _{A'} \prec \rho _A$ if the block
$A$ is a sub-block of $A'$ \cite{size}. Indeed, the majorization
relation follows from the fact that any two eigenvalue
distributions, corresponding to two different block sizes, only
cross \textit{once} when the eigenvalues $\lambda_i$ are plotted
versus the eigenvalue index $i$. That is, an index $i^*$ exists,
such that $\lambda_i$ decreases with increasing $\ell$ for $i\leq
i^*$, and $\lambda_i$ increases with increasing $\ell$ for $i>i^*$.
The uniqueness of this crossing is guaranteed by the fact that the
eigenvalue distribution follows $\lambda_i \propto q^{\beta_i}$ for
both bulk and boundary conformal theories, apart from degeneracies
$n_i$. Here $q=\exp(2\pi i \tau)$ for conformal theories and $q=\exp
(-\pi \delta)$ for boundary conformal theories, and $\beta_i$ are
integers related to the scaling dimensions of the descendant
operators. The conformal invariance requires that both $\tau$ and
$\delta$ should be proportional to $1/\ln \ell$. The discreteness of
$n_i$ and $\beta_i$ ensures that they do not change when the block
size $\ell$ is varied (for finite $L$, $\ell$ must be restricted to
even or odd values if the model has a parity effect).

Next, we consider the behavior of the R\'{e}nyi entropy along
boundary RG flows. By combining our CFT results with the
$g$-theorem, it follows that the R\'{e}nyi entropy decreases (more
precisely, does not increase) along boundary RG flows. In
particular, the largest eigenvalue $\lambda_1$ of $\rho_A$ increases
along the boundary RG flow. Furthermore, one may argue that along
the boundary RG flow, the eigenvalues of $\rho_A$ take the same form
$\lambda_i \varpropto q^{\beta_i}$ as at the conformally invariant
fixed point at the end of the flow, except that the dependence of
$q$ on $\ell$ is different (again, the discreteness of the
degeneracies $n_i$ and $\beta_i$ ensures that they remain the same
along the flow). However, for any nonzero $h_b$, the bulk
criticality requires that the dependence $\delta \sim 1/\ln \ell$ is
recovered for $\ell > \zeta$. It then follows that there is {\it one
and only one} crossing for the eigenvalue distributions versus $i$
along the boundary RG flows.   More precisely, an index $i^*$
exists, such that $\lambda_i$ increases for $i\leq i^*$, and
decreases for $i>i^*$ along the flows.  This in turn implies the
majorization relation.

\begin{figure}
 \centering
\includegraphics[width=7.55cm,height=6.0cm]{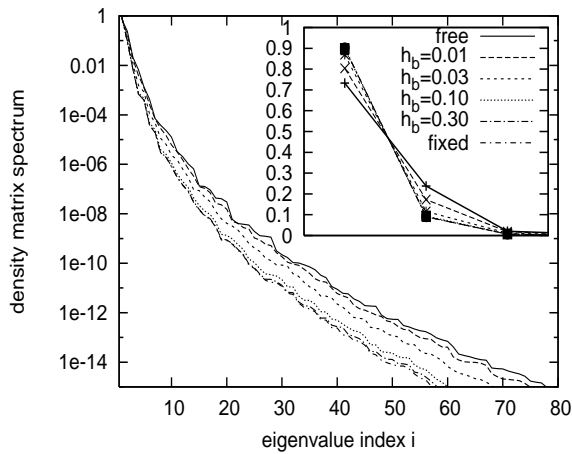}
\caption{\label{fig:IsingS2} Eigenvalue distributions for the
reduced density matrix of a block of size $\ell=128a$ in the bulk
critical transverse Ising model (\ref{eq:HamIsing}) for various
values of the boundary magnetic field $h_b$. The quasi-exact DMRG
results confirm the majorization relation along the boundary field
RG flow (see text). The inset shows that the crossing of the spectra
occurs at $i^*=1$.}
\end{figure}

A rough estimate for the crossing index $i^*$ can be obtained from
the CFT predictions for the largest eigenvalue $\lambda_1$. This
gives $i^* \sim \ell^{c/6}$ for PBC and $i^* \sim \ell^{c/12}$ for
OBC. Thus the block size $\ell$ must be sufficiently large to
observe that the crossing occurs at $i^* >1$. For instance, to see
that the second largest eigenvalue decreases with increasing $\ell$,
$\ell$ should at least be $1800a$ for the semi-infinite XX chain, as
estimated from the exact solution \cite{peschel99}. DMRG
calculations of the reduced density matrix spectra show majorization
along boundary RG flows, where eigenvalues down to $\sim 10^{-15}$
are considered. Fig.~\ref{fig:IsingS2} shows the crossing point
$i^*$ of the spectra in the bulk critical transverse Ising model
with block size $\ell=128a$ and various values of the boundary field
$h_b$. The crossing of the spectra occurs here at $i^*=1$.

For the Heisenberg XXX model ($\Delta=1$) the R\'{e}nyi entropy does
not depend on the boundary magnetic field: $\partial
S_{\alpha}/\partial h_b=0$. These constraints (of which there is an
infinite number, one for each $\alpha$) imply that the spectra of
$\rho_A$ do not vary with $h_b$. Therefore the presence of a line of
fixed points here amounts to the statement that all of these
critical points share the same eigenvalue distribution, so there is
no boundary RG flow (a boundary RG flow would require a preferential
direction characterized by (strict) majorization).

{\it Conclusions.} We have investigated the interrelations between
the boundary entropy, the Affleck-Ludwig $g$-theorem and
entanglement in quantum spin systems with boundary interactions. The
intrinsic irreversibility along boundary RG flows, as embodied in
Affleck-Ludwig $g$-theorem, is connected with the majorization
relation solely characterized by the ground state itself. The
results will bring new insights into our understanding of 2D
statistical mechanical systems with boundary interactions and
(perturbed) conformal theories defined on manifolds with boundaries.
The equivalence between 1D quantum systems and 2D classical systems
indicates that majorization relations should still be preserved in
2D classical systems, even if it does not make sense to speak of
quantum entanglement for classical cases.

During the preparation of this manuscript, two preprints~\cite{zhao}
appeared which reached the same conclusion as us regarding the
connection between the von Neumann entropy and the single-copy
entanglement for bulk conformal theories.  We thank Guifr\'{e} Vidal for
enlightening discussions and comments. HQZ and JOF thank the
Australian Research Council for support. TB thanks the DFG for support.


\begin{thebibliography}{99}

\bibitem{sachdev} S. Sachdev, {\it Quantum Phase Transitions}
(Cambridge University Press, 1999).

\bibitem{nielsen} T. J. Osborne and M. A. Nielsen, Phys. Rev. A
\textbf{66}, 032110 (2002); A. Osterloh \textit{et al.}, Nature
\textbf{416}, 608 (2002).

\bibitem{latorrevidal} G. Vidal \textit{et al.}, Phys. Rev. Lett.
\textbf{90}, 227902 (2003); J. I. Latorre \textit{et al.}, Quantum
Inf. Comput. \textbf{4}, 48 (2004).

\bibitem{korepin} V. E. Korepin, Phys. Rev. Lett. \textbf{92}, 096402
(2004); B.~Q. Jin and V. E. Korepin, J. Stat. Phys. \textbf{116}, 79
(2004).

\bibitem{levine} G. C. Levine, Phys. Rev. Lett. \textbf{93}, 266402
(2004); G. Refael and J. E. Moore, Phys. Rev. Lett. \textbf{93},
260602 (2004); J. P. Keating and F. Mezzadri, Phys. Rev. Lett.
\textbf{94}, 050501 (2005); M. B. Plenio \textit{et al.}, Phys. Rev.
Lett. \textbf{94}, 060503 (2005).

\bibitem{calabrese} P. Calabrese and J. Cardy, J. Stat. Mech.
P06002 (2004).

\bibitem{casini1} H. Casini and M. Huerta, Phys. Lett. B \textbf{600},
142 (2004).

\bibitem{casini2} H. Casini \textit{et al.}, J. Stat. Mech. P07007
(2005).

\bibitem{eisert} J. Eisert and M. Cramer, Phys. Rev. A \textbf{72},
042112 (2005).

\bibitem{bennett} C. H. Bennett \textit{et al.}, Phys. Rev. A \textbf{53},
2046 (1996).

\bibitem{wilczek} C. Holzhey \textit{et al.}, Nucl. Phys.
B \textbf{424}, 44 (1994).

\bibitem{latorre} J. I. Latorre \textit{et al.}, Phys. Rev. A
\textbf{71}, 034301 (2005).

\bibitem{zamolodchikov} A. B. Zamolodchikov, JETP Lett. \textbf{43},
731 (1986).

\bibitem{olkin} A. W. Marshall and I. Olkin, {\it Inequalities:
Theory of Majorization and its Applications} (Academic Press,
1979).

\bibitem{orus} R. Or\'{u}s, Phys. Rev. A \textbf{71}, 052327 (2005).

\bibitem{binder} K. Binder, in {\it Phase Transitions and Critical
Phenomena}, vol. 8, eds. C. Domb and J. Lebowitz, (Academic, London,
1983).

\bibitem{ghoshal} S. Ghoshal and A. B. Zamolodchikov, Int. J. Mod. Phys.
A \textbf{9}, 3841 (1994).

\bibitem{affleck} I. Affleck and A. W. W. Ludwig, Phys. Rev. Lett.
\textbf{67}, 161 (1991); D. Friedan and A. Konechny, Phys. Rev.
Lett. \textbf{93}, 030402 (2004).

\bibitem{preskill} J. Preskill, J. Mod. Opt. \textbf{47}, 127
(2000).

\bibitem{White1992-11}
 S. R. White, Phys. Rev. Lett. \textbf{69}, 2863 (1992).

\bibitem{Schollwoeck2005}
  U. Schollw\"{o}ck, Rev. Mod. Phys. \textbf{77}, 259 (2005).

\bibitem{cardy-lewellen} J. Cardy,  Nucl. Phys. B \textbf{324}, 581
(1989); J. Cardy and D. Lewellen, Phys. Lett. B \textbf{259}, 274
(1991).

\bibitem{difrancesco} P. Di Francesco, P. Mathieu, and D.
S\'{e}n\'{e}chal, {\it Conformal Field Theory}, (Springer, Berlin,
1997).

\bibitem{bs} The boundary states are also well-defined for systems
off criticality, see Ref.~\cite{ghoshal}.

\bibitem{uni} For a composite system $AB$ consisting of two
non-interacting systems $A$ and $B$, we have
$S_{\alpha}^{{AB}}(c^{{AB}},\ell/a)=S_{\alpha}^{{A}}
(c^{{A}},\ell/a)+S_\alpha^{{B}}(c^{{B}},\ell/a)$ and
$c^{{AB}} =c^{{A}}+c^{{B}}$. This implies that
$S_{\alpha}^{{A}}(c^{{A}},\ell/a)=c^{{A}}f_{\alpha}(\ell/a) + h^{{A}}_{\alpha}$. So
$f_{\alpha}$ is not model-specific whereas $h_{\alpha}$ is.

\bibitem{aff98} I. Affleck, J. Phys. A: Math. Gen. \textbf{31}, 2761 (1998).


\bibitem{callan} C. G. Callan \textit{et al.}, Nucl. Phys.
\textbf{B} 422, 417 (1994).

\bibitem{cardy05} J. Cardy, in {\it Encyclopedia of Mathematical
Physics}, eds. J.-P. Fran\c{c}oise \textit{et al.}, (Elsevier,
2005).

\bibitem{Verstraete2004-4}
 F. Verstraete \textit{et al.}, Phys. Rev. Lett. \textbf{93}, {227205}
 (2004).

\bibitem{schur} A function $\phi$  is Schur-concave if $\lambda \prec \mu
\Rightarrow \phi (\lambda) \geq \phi (\mu)$.

\bibitem{size} A rigorous proof for this fact was attempted in Ref.~\cite{orus}
for bulk conformal theories, where it was claimed that all
eigenvalues except the largest one increases with increasing block
size $\ell$. This is not valid, as one can see from the exact
solution for the XX model \cite{peschel99}, which shows that the
second largest eigenvalue first increases with $\ell$ before it
starts decreasing. The turning point is at $\ell/a\sim 1800$.

\bibitem{peschel99} I. Peschel \textit{et al.}, Ann. Physik (Leipzig) \textbf{8}, 153
(1999).

\bibitem{zhao} I. Peschel and J. Zhao, quant-ph/0509002; R. Or\'{u}s \textit{et al.}, quant-ph/0509023.



\end{thebibliography}
\end{document}